# Minireview on Disordered Optical Metasurfaces

Philippe Lalanne[1,*], Alexandre Dmitriev,[2] Carsten Rockstuhl,[3] Alexander Sprafke,[4] and Kevin Vynck[5]

[1]Laboratoire Photonique Numérique et Nanosciences (LP2N), Université de Bordeaux, Institut d'Optique Graduate School, CNRS, 33400 Talence, France
[2]Department of Physics, University of Gothenburg, 41296 Gothenburg, Sweden
[3]Institute of Theoretical Solid State Physics, Karlsruhe Institute of Technology, 76131 Karlsruhe, Germany
[4]Institute of Physics, Martin Luther University Halle-Wittenberg, 06120 Halle, Germany
[5]Institut Lumière Matière (iLM), Université Claude Bernard Lyon 1, CNRS, 69100 Villeurbanne, France

[*]philippe.lalanne@institutoptique.fr

**Abstract**. The use of coherent wave phenomena to enhance device performance is a cornerstone of modern optics. In juxtaposition to (locally) periodic metasurfaces, their disordered counterparts exhibit an interplay of destructive and constructive interferences occurring at the same spatial and spectral frequencies. This attribute provides disordered metasurfaces with a remarkable degree of flexibility, setting them apart from the constraints of periodic arrangements. Hereafter, we provide a concise overview of the cutting-edge developments and offer insights into the forthcoming research in this dynamic field.

## 1. Introduction

Disorder has often been viewed as a constraining element in nanophotonics, which might explain the limited coverage of the subject of disordered metasurfaces in the existing literature [1,2]. Nevertheless, there exists a plenty of reasons to delve into this area. Let us highlight just a few. The presence of disorder is ubiquitous across manifold complex structures. Leveraging disordered metasurfaces, as opposed to their meticulously ordered counterparts, holds promise in realizing extensive, cost-effective devices. It also paves the way to a wealth of specialized applications, such as anti-reflection coatings [3], augmented absorption [4], surface-enhanced Raman scattering [5], alongside contemporary applications to be discussed below.

This article aims to survey the current state of research and offer a forward-looking perspective on the topic. We highlight a few promising future applications and anticipate forthcoming advancements in design and fabrication.

## 2. State-of-the-art

The general problem of light scattering by disordered metasurfaces made of resonant nano-objects (a.k.a. meta-atoms) is very complicated due to multiscale effects, namely the multipolar resonances of the individual particles, their interaction with the substrate, and their mutual interaction [6,7]. In this section, we start by outlining some fundamental concepts on coherent vs incoherent scattering. We then familiarize the reader with computational tools and approximate models that predict the scattering properties of disordered metasurfaces, and finally examine recent advances in fabrication techniques.

### Diffuse (incoherent) vs specular (coherent) light

Within the context of wave scattering by rough surfaces – a long-standing topic with many applications [8,9,10] –, the distinction between diffuse (incoherent) light and specular (coherent) light emerges as a pivotal consideration. These two components have different physical origins, as we will see below, and have significant implications for optical effects – for instance, their relative weight play a decisive role in determining the glossiness or matteness of a surface.

This distinction can be understood from the statistical properties of the scattered light. In this framework, one considers separately the average and fluctuating values of physical parameters, with the averages computed among a large statistical ensemble of independent realizations. The quantities of main interest in the present case are electromagnetic field characteristics, including the field itself, its intensity or root mean square, along with auxiliary quantities such as the Poynting vector and the energy density that a wave carries.

Formally, the electric field $\mathbf{E}_S$ scattered by a disordered system can be expressed as the sum of its ensemble average (the specular light), calculated among many disorder realizations, and a fluctuating term (the diffuse light) that varies from configuration to configuration,

$$\mathbf{E}_S = \langle \mathbf{E}_S \rangle + \delta \mathbf{E}_S, \tag{1}$$

with $\langle \delta \mathbf{E}_S \rangle = \mathbf{0}$.

Figure 1 illustrates this decomposition using fully-vectorial simulations conducted on a finite metasurface composed of parallel infinitely-long silicon nanocylinders in air. The metasurface is illuminated at a 30° angle of incidence by a plane wave polarized parallel to the cylinder $y$-axis. In the rightmost panel, the fluctuating field $\delta \mathbf{E}_s$ showcases a complex pattern known as speckle [11], which arises from the intricate interplay of near-field interactions among many monochromatic waves scattered by multiple scatterers. In contrast, the central panel presents the average field $\langle \mathbf{E}_s \rangle$ that readily takes the form of a directional reflected light beam indicated by black arrows. This average field describes the diffracted light from a homogenized metasurface characterized by effective parameters that vary non-uniformly in both the $x$ and $z$ directions due to boundary effects.

Given $\langle \delta \mathbf{E}_s \rangle = \mathbf{0}$, we can readily demonstrate that the averaged intensity $\langle |\mathbf{E}_s|^2 \rangle$ of the light scattered by the metasurface can be decomposed into two components,

$$\langle |\mathbf{E}_s|^2 \rangle = |\langle \mathbf{E}_s \rangle|^2 + \langle |\delta \mathbf{E}_s|^2 \rangle, \qquad (2)$$

in which we identify the squared norm of the average field, which corresponds to the specular intensity (first term), and the average of the squared norm of the fluctuating field, i.e. the diffuse intensity or the averaged speckle intensity (second term).

The properties of statistically translationally-invariant (i.e., infinite) disordered metasurfaces are fully determined by the Bidirectional Scattering Distribution Function (BSDF), a radiometric quantity introduced in the 1960s, which describes how surfaces scatter light for all possible planewave illuminations. Formally, the BSDF relates an incoming irradiance $E_i$ (in $W \cdot m^{-2}$) to a scattered radiance $L_s$ (in $W \cdot m^{-2} \cdot sr^{-1}$),

$$\text{BSDF}(\hat{\mathbf{k}}_s, \hat{\mathbf{e}}_s, \hat{\mathbf{k}}_i, \hat{\mathbf{e}}_i, \omega) = \frac{dL_s(\hat{\mathbf{k}}_s, \hat{\mathbf{e}}_s, \omega)}{dE_i(\hat{\mathbf{k}}_i, \hat{\mathbf{e}}_i, \omega)}, \qquad (3)$$

where $\omega$ is the frequency, $\hat{\mathbf{k}}_i$ and $\hat{\mathbf{k}}_s$ are the (unit) wavevectors of the incident and scattered planewaves, and $\hat{\mathbf{e}}_i$ and $\hat{\mathbf{e}}_s$ are the polarization (unit) vectors of those waves. Following the previous considerations, the BSDF can then be conveniently decomposed into two terms, $\text{BSDF} = \text{BSDF}_{\text{spec}} + \text{BSDF}_{\text{diff}}$, the specular and diffuse components.

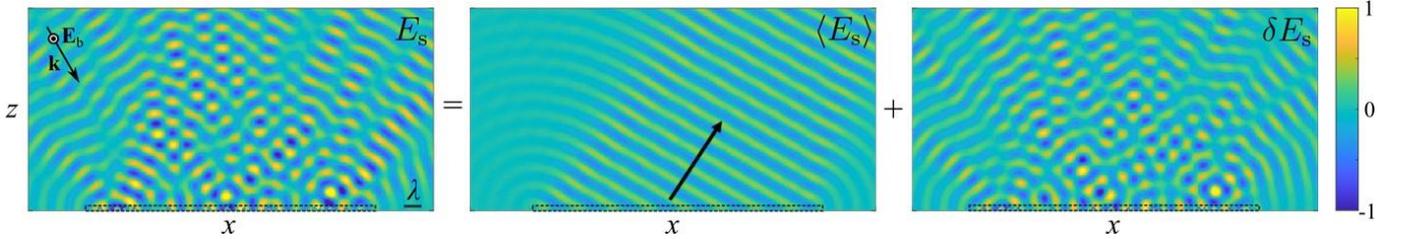

**Fig. 1.** Illustration of the average (coherent) and fluctuating (incoherent) components of the scattered field (Eq. 1) with fully-vectorial simulations for a finite-size disordered metasurface composed of identical Si nanocylinders in air. The incident (background) field $\mathbf{E}_b$ is a planewave ($\lambda = 440$ nm) polarized along the $y$-direction. It is incident from the top with an angle of 30° with respect to the surface normal. The rods are parallel to the $y$-axis and are placed using a random sequential addition algorithm with a non-overlapping condition. The 2D simulations are made for assemblies of 15 infinitely-long nanocylinders (140 nm diameter, $n_{\text{Si@440nm}} = 4.793 + 0.109i$), resulting in a side length of 7.0 μm ($\approx 16\lambda$) at a surface coverage of 30%. Statistical averaging are collected with 1000 independent disorder configurations. The $y$-component of the scattered field (component parallel to the cylinder $y$-axis), $E_s$, is normalized to the amplitude of the background field $|\mathbf{E}_b|$.

**Full-wave analysis**
Predicting the BSDF of statistically translationally-invariant metasurfaces featuring resonant particles on layered substrates stands as an outstanding challenge [12,13]. Two complementary strategies are employed. The first approach involves solving Maxwell equations with utmost precision, while the second approach (next subsection) relies on approximate models.

In the former approach, the strategy is twofold: first, solving Maxwell equations for larger and larger metasurfaces, and second, inferring the BSDF of infinitely large metasurfaces by extrapolation. In contrast to traditional general-purpose solvers such as the Finite-Difference Time-Domain method, disordered metasurfaces benefit significantly from specialized numerical techniques. The latter harness the unique morphologies of these structures and enhance their efficiency by precomputing light scattering by the individual scatterers that compose the metasurface.

In the widely used T-matrix method, the field outside the scatterer is expanded using spherical vectorial wave harmonics and a T-matrix is utilized to describe how each scatterer transforms incident into scattered multipolar fields. The method has been extensively explored over decades [14] and is now available through multiple open-source implementations [14,15]. Alternative methods account for multiple scattering using vectorial Green functions and treat scatterers as ensembles of equivalent electric and magnetic surface currents [16,17], or employ numerical non-local dipoles [18] obtained by numerically solving an inverse scattering problem. A detailed comparative analysis of these methods strengths and limitations is beyond this roadmap's scope and is deferred to a comprehensive future review [19].

Two approaches exist for the extrapolation step. In the field stitching approach, Maxwell equations are solved for numerous small square subdomains, each corresponding to an independent realization. These subdomains are then stitched together to artificially expand the metasurface area, with field discontinuities smoothed at subdomain boundaries to mitigate artifacts [20]. In the supercell approach, identical subdomains are 'stitched' with pseudo-periodic boundary conditions. As the artificial period increases, both approaches generally yield stable numerical outcomes for specular and diffuse light [21,22]. Although finite-size computations unavoidably introduce boundary effects [22], the observed relative errors are often encouragingly below 10%.

However, the two approaches necessitate numerous computations to assess statistical convergence, and they offer limited physical insights. The challenge of effectively simulating infinite metasurfaces remains, highlighting the need for approximate models.

**Approximate models**

The dominant approach for modeling the BSDF assumes that each constituent particle scatters light independently, a simplification known as the independent scattering approximation (ISA) in the physics of waves in complex media. While this approximation that ignores multiple scattering might seem rudimentary to readers, it is important to recognize that the initial stages of designing ordered metasurfaces also lean on a similar premise. Indeed, lookup tables are typically derived from isolated metaatoms or periodic arrays.

The strength of the ISA lies in its ability to highlight the underlying physics and facilitate inverse design processes. For a monolayer comprising $N$ identical particles randomly positioned at coordinates $\mathbf{r}_p = [x_p, y_p, z_p = 0], p = 1 \ldots N$, the ISA simply leads to

$$\langle I_s \rangle (\hat{\mathbf{k}}_s, \hat{\mathbf{e}}_s, \hat{\mathbf{k}}_i, \hat{\mathbf{e}}_i, \omega) = I_0 N \frac{d\sigma_s}{d\Omega}(\hat{\mathbf{k}}_s, \hat{\mathbf{e}}_s, \hat{\mathbf{k}}_i, \hat{\mathbf{e}}_i, \omega) S(\mathbf{q}), \tag{4}$$

for the total radiant intensity $\langle I_s \rangle$ (measured in $\text{W} \cdot \text{sr}^{-1}$), encompassing both coherent and incoherent components, scattered by the particles. In Eq. (4), $\frac{d\sigma_s}{d\Omega}$ (measured in $\text{m}^2 \cdot \text{sr}^{-1}$) is the differential scattering cross section of individual particles – a quantity directly related to the so-called form factor in condensed matter physics. $I_0$ is the incident intensity (measured in $\text{W} \cdot \text{m}^{-2}$). $S(\mathbf{q})$ is the structure factor, which incorporates the effect of far-field interference between pairs of particles on the scattered intensity, and is defined as

$$S(\mathbf{q}) = 1 + \frac{1}{N} \langle \sum_{m \neq n}^{N} \exp[i \mathbf{q} \cdot (\mathbf{r}_m - \mathbf{r}_n)] \rangle, \tag{5}$$

where $\mathbf{q} = \frac{\omega}{c} n_b (\hat{\mathbf{k}}_i - \hat{\mathbf{k}}_s)$ is the wavevector difference between the incident and scattered waves and $n_b$ is the refractive index of the background medium.

The structure factor can be readily evaluated through statistical averaging. On the other hand, accurate determination of the form factor involves iteratively solving Maxwell equations for multiple wavelengths, incidences, and polarizations, followed by applying a near-to-far-field transformation to obtain the radiation pattern. Alternatively, a quicker approach with deeper insights can be achieved by expanding the differential scattering cross-section using the resonance modes of the nanoparticles [23].

Exploiting Eq. (4), one can derive straightforward expressions for the BSDF of infinite metasurfaces with $N \to \infty$. However, due to the inherent limitations of the ISA, the model's predictions exhibit notable inaccuracies for high particle densities, grazing incident angles, or large scattered angles. These inaccuracies prompt the need for further refinements and extensions. For readers seeking a more comprehensive understanding, let us note that advanced models for the diffuse contribution in the BSDF have been presented in [21]. For the specular contribution, an overview of various models can be found in a recent book chapter [24] and the references therein. Noteworthy is the remarkable efficacy demonstrated by the quasi-crystalline approximation [25,26].

**Fabrication**

Several techniques exist for fabricating disordered metasurfaces. Top-down patterning methods, which make use of electron-beam or optical lithography, facilitate the precise positioning of the meta-atoms with nanometer-level precision and a fine-tuned control over meta-atom size and shape. Furthermore, with the advent of nanoimprint lithography [27] and soft lithography [28], they have opened up new avenues for cost-effective fabrication.

Conversely, bottom-up approaches prioritize scalability and cost-efficiency over precision. Although they may not match top-down techniques in shaping and arranging nano-objects, their utility remains unequalled.

Among metasurfaces fabricated by scalable and cost-effective techniques, semicontinuous colloidal films composed of metal nano-islands stand out. These films naturally arise from deposition techniques that interrupt metal film growth before percolation threshold is reached [29]. Originally investigated for enhanced Raman scattering [30], these ultra-affordable films can be mass-produced by the glass industry on extensive scales, spanning tens of square meters. Because island dimensions and interspace distances are much smaller than visible wavelengths, nanoisland metasurfaces act as coherent layers, absorbing rather than scattering incident light. Alternatives, such as utilizing block copolymer templates for self-assembly [31], also facilitate monolayer creation with such tiny particles.

Producing disordered configurations with larger particles involves assembling pre-synthesized particles onto surfaces. Various techniques fit within this category:

- Colloidal lithography: This method hinges on nearly uniform in size colloids in solutions of materials like polystyrene, PMMA, or Si nanospheres [32] (Fig. 2a-b). After drop-casting and drying, a monolayer forms [33], subsequently serving as a mask to imprint patterns into underlying substrates. Since the sizing of polymer beads and strategies for self-assembly — utilizing repulsive and attractive forces for instance — are well-mastered (Fig. 2b), the transferred patterns, often arrays of nanodiscs or nanoholes in various materials, are fabricated on large scales with high fidelity and few aggregates.
- Self-assembly: With this method, colloidal nanoparticles, initially stabilized in solution, arrange on surfaces via methods like drop-casting and dip-coating [34]. Diverse particle shapes, sizes and materials are commercially available, and controlled short-range arrangements can be achieved through nanoparticle and substrate surface functionalization (Fig. 2c).

For deeper insights, recent comprehensive reviews provide substantial information on the subject [35,36,37,38,39].

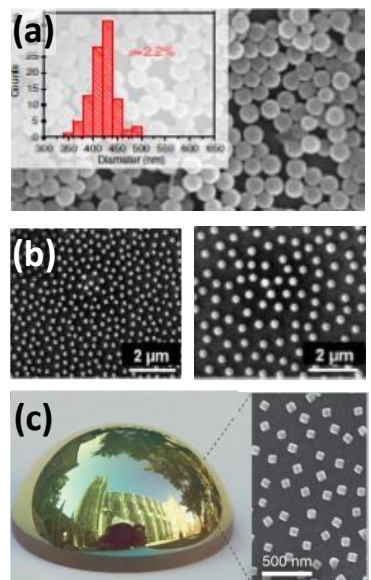

**Fig. 2** A few selected examples of disordered metasurfaces fabricated with colloidal chemistry. (**a**) Availability of almost monodisperse nanospheres in various dielectric materials, e.g. PMMA, high-index Si nanospheres in the present figure [32]. (**b**) Amorphous colloidal monolayers of polystyrene nanospheres with sizes of 170 and 270 nm with short-range order [33]. (**c**) Monolayer of Au nanocubes with correlated disorder on a 2.5 cm-diameter half sphere [34].

## 3. Future directions and outlook

Disordered optical metasurfaces have unveiled a plethora of applications, some of which can be traced back to historical contributions, such as porous film-based anti-reflective coatings by Fraunhofer in 1887 [3], and observations of ultra-strong Raman signals from semicontinuous nanoisland metal films in the 1970s [5]. Over the past two decades, the field of disordered optical metasurfaces has experienced remarkable activity, encompassing a diverse array of applications. These include wide-angle broadband absorption [40,34f,41], anti-counterfeiting measures [42], anti-reflection coating [3f,43,20f], wavefront manipulation [44,45,46], transparent displays [47,48], chiral films [49], fade-resistant structural coloring [50,51,52,41f,53,54], impressive visual effects [21f,23f], plasmonic electronic paper [55], solar radiation management in architectural and consumer products glazing [56,57,58,59], light trapping in solar cells [33f], and light extraction in LEDs [60,61,62,63].

As nanofabrication capabilities continue to advance, these applications are progressively moving closer to industrial-scale production. In the subsequent sections, we delve into emerging applications that are poised to inspire numerous studies and emphasize two prospective avenues for enhancing manufacturing flexibility and reducing manufacturing costs.

**Prospective applications**

Anticipating the forthcoming wave of research, we expect a concerted effort to tackle the enduring question of how to master the nanopatterning of surfaces for shaping the far-field radiation across both spatial and spectral frequency domains. The ultimate ambition is to attain comprehensive command over both specular and diffuse light. It encompasses manipulating independently these two components, sculpting the directional patterns of diffuse light, monitoring alterations in light color for varying viewing and illumination angles, and, in essence, molding the BSDF at will (Fig. 3a).

Two key physical parameters that play pivotal roles can be used for design (see Eq. 4). Through the form factor, an assortment of shapes and high-index materials may be used to manipulate the resonances of the meta-atoms at the nanoscale. These resonances may be further enriched through mode hybridization at the wavelength scale to achieve wavelength-selective control over the directionality of scattered light [64,65,66]. Through the structure factor, we may engineer intricate mesoscale interferences. This engineering encompasses a vast landscape of arrangements [1f,67], ranging from full periodicity to entirely uncorrelated disorders. This gamut includes quasicrystals, weakly defective crystals, hyperuniform configurations, and more. For example, enforcing greater inter-particle distances engenders a short-range amorphous-type correlation that mitigates the presence of diffuse light around the specular direction [68,21f], regardless of the angle of incidence.

This effect is even more pronounced within stealthy hyperuniform metasurfaces, which entirely suppress scattering within an angular range centered on the specular direction [1f]. Ongoing endeavors in the domain of hyperuniform disordered scattering structures, particularly for solar cells, are only just emerging [68,69,70,71]. A particularly remarkable recent achievement is the 2.5-fold enhancement in absorption of a 1µm-thick silicon slab, thanks to a hyperuniform metasurface (Fig. 3b) meticulously designed to trap incident light into the slab guided modes [71f].

Disordered optical metasurfaces also hold significant promise for applications in applied and fine arts. Metasurfaces offer the potential to manipulate the visual characteristics—such as color, glossiness, haze, transparency, and iridescence—of macroscopic objects of any conceivable shape [21f]. A recent breakthrough in this domain introduced a multiscale numerical tool capable of generating lifelike images of diverse macroscopic objects covered with disordered metasurfaces (Fig. 3c). This tool not only forecasts subtle effects resulting from the interplay between diffuse and specular light but also unveils entirely new visual effects.

Future exploratory directions may capitalize on the monolithic integration of multiple metasurface layers. A compelling feature of these surfaces is their distinct behavior concerning diffuse light in transmission versus reflection. This contrast can be readily comprehended by examining the structure factor, $S(\boldsymbol{q})$ in Eq. (5), in which $\boldsymbol{q}$ signifies the difference in wavevector between the scattered direction and incident directions. It becomes evident that $(\mathbf{k}_i - \mathbf{k}_s) \cdot (\mathbf{r}_m - \mathbf{r}_n)$ in the exponential term remains independent of the normal coordinate $z$ around the specular transmitted direction (where $\mathbf{k}_s \equiv \mathbf{k}_t \approx \mathbf{k}_i$). In contrast, this term does vary with the normal coordinate around the specular reflected direction (where $\mathbf{k}_s \equiv \mathbf{k}_r \approx -\mathbf{k}_i$). This intrinsic property has recently been ingeniously harnessed to create a transparent display [48] that selectively scatters reflected light while preserving transmitted light integrity (Fig. 3d).

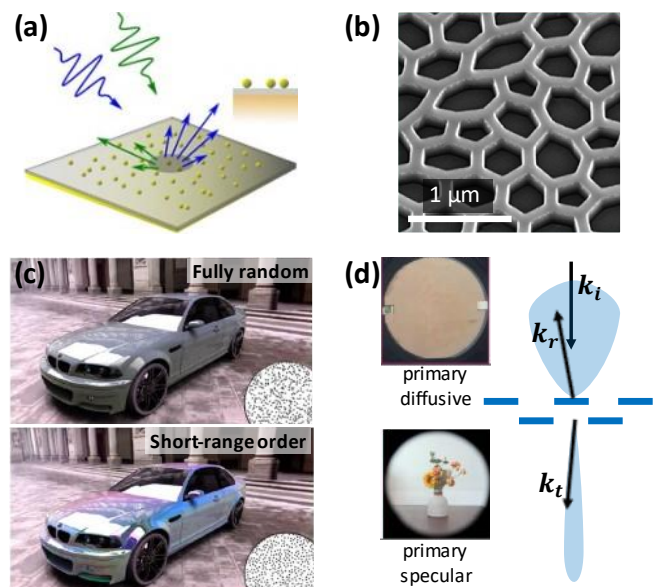

**Fig. 3** (**a**) Harnessing the BRDF towards full manipulation of diffuse and specular light. In the sketchy example, the surface color varies abruptly, differently from the continuous color variation of classical iridescence, when the light source and/or the observer position move. (**b**) Hyperuniform metasurface for light trapping in solar cell application. After [71]. (**c**) Rendered images showing the visual appearances of a car coated with disordered arrays of identical Ag nanospheres with a density of 4 µm$^{-1}$. Top: no structural correlation. Bottom: short-range correlated disorder. Adapted from [21]. (**d**) Transparent display made of a bilayer metasurface which offer completely different properties in reflection and transmission. The rightmost images are camera pictures of a flower bouquet taken in reflection (top) or in transmission (bottom). After [48].

## Metasurfaces on curved substrates

An overwhelming number of nanopatterning methods, including optical and electronic lithography, nanoimprint lithography [27f], soft-lithography [28f], DNA-templating [72], functional block copolymers [73], are available nowadays. They often function on 2D and atomically flat surfaces.

In general, adapting methods tailored for flat surfaces to curved ones necessitates significant changes to mitigate curvature-induced variations in feature attributes like size, orientation, and spacing. While some fabrication methods can accommodate relatively simple curvatures, the adaptation challenge grows as curvature intensifies. This puts severe limitations onto the use of metasurfaces in a variety of every-day-use or advanced technologies, where materials with non-planar, rough, soft, mechanically agile or chemically sensitive surfaces are used [74,75,76].

Recently, several transfer protocols have been proposed to overcome the limitations [77,34f,76f,78]. Figure 4a shows the example of a pyramid-textured Si surfaces which has been patterned with air holes to lower reflection at the Si-air interfaces and enhance the total absorption. The hole pattern is obtained from a flat parent substrate template manufactured with colloidal lithography (other bottom-up or top-down approaches may be used as well). The key is the use of a 10 nm-thin carbon film deposited on a sacrificial layer for the template. Such a thin film combines extreme strength with a large degree of flexibility that facilitates conformation of the transferred pattern to nearly any macroscopic or microscopic 3D object.

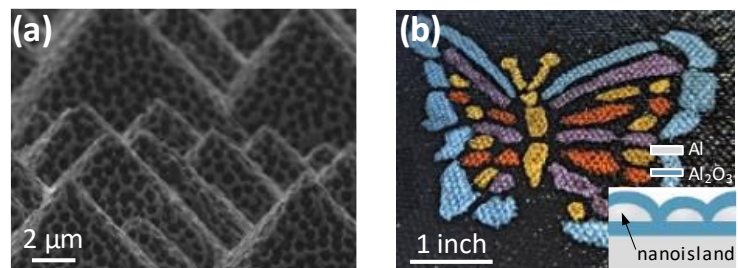

**Fig. 4** (**a**) Metasurfaces fabricated on curved substrates: air holes etched in pyramid-textured Si surfaces. Adapted from [Mas16] (**b**) Production of vivid colors despite a strong polydispersity in the size and shape of Al nanoislands used to resonantly absorb light. The resonance is due to nanogap plasmons supported by the capped nanoislands obtained by dewetting on $Al_2O_3$/Al substrate (inset). The photography shows a black canvas painted with oil–based plasmonic paints obtained by dispersing metasurface flakes in commercial paint oils. Adapted from [Cen03].

## Striving for ultra-low costs with augmented resilience to fabrication imperfections

Optical metasurfaces, which harness resonant nanostructures at subwavelength scales, impose strong constraints on the nanofabrication precision. These constraints are growing in significance, as metasurfaces progress towards practical applications that require large sample areas. In this evolving landscape, the emergence of metasurfaces designed to exhibit resilience against imperfections becomes increasingly crucial.

Recent studies have converged to provide compelling evidence that disordered arrays of nanoparticles, when placed on a transparent dielectric spacer above a metallic substrate, exhibit an extraordinary resilience to variations in material, size, and nanoparticle shape.

One fundamental mechanism that imparts resilience is through the process of mode hybridization [23f]. In scenarios where plasmonic nanoparticles are situated on low-index substrates without undergoing hybridization, the electromagnetic field becomes profoundly confined around the individual particle. Consequently, even slight variations in material composition or shape induce significant shifts in resonance frequencies. For context, even a minor alteration of 10 nm in the dimensions of a metal nanoparticle leads to a resonance wavelength shift of 10-20 nm. In contrast, within hybridized modes, the field localization diminishes as it extends into the transparent layer. Consequently, hybridization inherently mitigates sensitivity to size and material polydispersity. This general attribute of hybrid modes has facilitated the discovery of a remarkable diffuse iridescence phenomenon, in which a mere two distinct and vivid colors are discernible across all angles of illumination and viewing, even when confronted with significant variations in nanoparticle size and shape across the metasurface [23f].

An alternative resilience mechanism emerges when nanoparticles are substantially smaller than the wavelength. This scenario is exemplified by semicontinuous nanoisland films on a dielectric spacer atop a metal substrate, where empirical observations reveal that the pivotal factor shaping the specular reflectance spectrum is the spacer thickness [51], rather than the typical expectation of the nanoparticle material choice. This material-independent response has been experimentally verified for various materials like Ag, Au, Pt, Pd, and numerically for Al, Cr, Cu, Ni, among others, under normal incidence [51]. Thus commonly accessible materials such as Cu or Al could substantially mitigate costs related to nanostructure-based coloration.

This striking property can be attributed to the amalgamation of localized plasmonic resonances at the nanoisland scale, hot spots within the interstitial gaps, and delocalized modes, irrespective of the metal used for the nanoislands.

This synergy results in broad absorption, further enhanced by the diversity in island size and shape. Consequently, the film uniformly absorbs the incident light, except for the Fabry-Perot resonance of the thin transparent spacer, for which the field is nearly null within the semicontinuous film.

Yet another illustration of resilience against manufacturing defects is observed in a recent work on ultralight plasmonic structural-color paints [54] with a notably low-cost morphology shown in the inset of Fig. 4b. Impressively, vibrant colors are attainable despite the substantial variance in nanoisland sizes. While plasmon hybridization leading to a confined gap-plasmon resonance within the alumina film stands as a logical explanation, it is important to acknowledge the influence of other mechanisms as well. Among these, the generation of subtractive color through the absorption of a "homogenized" film emerges as a noteworthy consideration, deviating from the diffusion-based coloration observed with larger nanoparticles.

## Acknowledgments

PL and KV acknowledge financial support from the french National Agency for Research (ANR) under the project "NANO-APPEARANCE" (ANR-19-CE09-0014). A. D. acknowledges the Swedish Research Council for Sustainable Development (Formas) (Project No. 2021-01390). PL acknowledges financial support from the European Research Council Advanced grant (Project UNSEEN No. 101097856).